\documentclass[aps,pre,preprint,superscriptaddress,showpacs]{revtex4-1}

\usepackage{amsmath, amsfonts, amsthm, amssymb, mathrsfs, enumerate, graphicx, cases, hyperref}
\usepackage[T1]{fontenc}
\usepackage[utf8]{inputenc}

\newcommand{\R}{{\mathbb R}}

\begin{document}


\title{Bondi mass with a cosmological constant}


\author{Vee-Liem Saw}
\email[]{Vee-Liem@ntu.edu.sg}
\affiliation{Division of Physics and Applied Physics, School of Physical and Mathematical Sciences, Nanyang Technological University, Singapore}


\date{\today}

\begin{abstract}
The mass loss of an isolated gravitating system due to energy carried away by gravitational waves with a cosmological constant $\Lambda\in\R$ was recently worked out, using the Newman-Penrose-Unti approach. In that same article, an expression for the Bondi mass of the isolated system, $M_\Lambda$, for the $\Lambda>0$ case was proposed. The stipulated mass $M_\Lambda$ would ensure that in the absence of any incoming gravitational radiation from elsewhere, the emitted gravitational waves must carry away a positive-definite energy. That suggested quantity however, introduced a $\Lambda$-correction term to the Bondi mass $M_B$ (where $M_B$ is the usual Bondi mass for asymptotically flat spacetimes) which would involve not just information on the state of the system at that moment, but ostensibly also its past history. In this paper, we derive the identical mass-loss equation using an integral formula on a hypersurface formulated by Frauendiener based on the Nester-Witten identity, and argue that one may adopt a generalisation of the Bondi mass with $\Lambda\in\R$ \emph{without any correction}, viz. $M_\Lambda=M_B$ for any $\Lambda\in\R$. Furthermore with $M_\Lambda=M_B$, we show that for \emph{purely quadrupole gravitational waves} given off by the isolated system (i.e. when the ``Bondi news'' $\sigma^o$ comprises only the $l=2$ components of the ``spherical harmonics with spin-weight 2''), the energy carried away is \emph{manifestly positive-definite} for the $\Lambda>0$ case. For a general $\sigma^o$ having higher multipole moments, this perspicuous property in the $\Lambda>0$ case still holds if those $l>2$ contributions are weak --- more precisely, if they satisfy any of the inequalities given in this paper.
\end{abstract}


\maketitle



\section{Introduction}\label{introduction}

The past couple of years have indubitably been a fascinating time to be working on gravitational physics. Since the inaugural direct detection of gravitational waves originating from a binary black hole system was announced in February 2016 (one century after Albert Einstein formulated his theory of general relativity in November 1915) \cite{LIGO}, several more such detections have been announced \cite{LIGO2,LIGO3,LIGO4,LIGO5}. Intriguingly, the most recent ones are not just signals picked up by LIGO's twin detectors, but also with VIRGO \cite{LIGO4,LIGO5} as well as other non-gravitational detections: including those coming from a binary neutron star system \cite{LIGO5}. It is certainly apt that this year's Nobel Prize in Physics goes to the scientists that represent the global collaboration to make direct observations of gravitational waves a reality, opening up exciting new territories for learning about our Universe.

Whilst it was considerably long before the experimentalists' breakthrough occurred, the amount of time taken to concretely build the \emph{theoretical foundation} for gravitational waves was of a comparable order of magnitude, when it was only firmly established by Bondi and his coworkers in the 1960s \cite{Bondi62,Sachs62}. Since then, we have better understood the global and asymptotic structure of spacetime. In particular, for an isolated system such that spacetime is asymptotically empty, there exists asymptotic symmetries and these asymptotic symmetries can be used in a Hamiltonian framework to obtain the Bondi mass \cite{Sachsss,BLY,200years} --- in agreement with Bondi's original proposal. There is an underlying assumption in the vast majority of the work back then, however: that spacetime is \emph{asymptotically flat} \footnote{There must be numerous exceptions, of course. For example, Penrose's conformal treatment of spacetime does include a cosmological constant \cite{Pen65}. Also, Hawking studied gravitational radiation in an expanding universe, and in some setup which reduced to the Bondi mass for asymptotically flat spacetimes \cite{Haw68}.}.

If we look back at the preceding Nobel Prize in Physics being awarded to the field of gravitation and cosmology, that was in 2011 for the discovery of the accelerated expansion of our Universe. That revelation meant that one cannot assume that the cosmological constant $\Lambda$ in Einstein's theory of general relativity is zero, without introducing extra properties to spacetime. Well, a way to account for this phenomenon is to simply admit $\Lambda>0$ into Einstein's theory. In fact, there have been investigations taking place over the past few years by many researchers employing a raft of methods to extend the notion of the Bondi mass and the energy carried by gravitational waves to include a positive cosmological constant. (See for instance, Ref. \cite{Vee2017d} for a review on the progress thitherto and the references therein.)

One approach to include a cosmological constant with any sign or value, i.e. $\Lambda\in\R$, is to use the Newman-Penrose formalism \cite{Vee2016,Vee2017,newpen62} \footnote{Ref. \cite{Vee2017} is an extension of Ref. \cite{Vee2016} to include Maxwell fields. The case with purely Maxwell fields is clear: the mass-loss formula for electromagnetic (EM) radiation says that outgoing EM waves strictly decrease the total mass-energy of the source, whilst incoming EM waves from elsewhere strictly increase its total mass-energy. The gravitational case, on the other hand, is not as straightforward. See also Ref. \cite{Vee2017b} for a concise summary of the results found in Refs. \cite{Vee2016,Vee2017}.}, as Newman and Unti also obtained the same mass-loss formula for an isolated gravitating system due to energy carried away by gravitational waves for $\Lambda=0$ \cite{newunti62} (during the same period when Bondi reported his results). The corresponding mass-loss formula with $\Lambda\in\R$ was derived in Ref. \cite{Vee2016} using this Newman-Penrose-Unti approach, with a proposal for the Bondi mass in a universe with $\Lambda>0$ being \footnote{Note that the term involving the $u$-derivative of the area element appearing in Eq. (126) in Ref. \cite{Vee2016} is zero, as pointed out in footnote ``s'' of Ref. \cite{Vee2017d}. We give an explicit proof in Appendix \ref{appendix}. Incidentally, whilst the ansatz for the fall-offs of various quantities used in Ref. \cite{Vee2016} were guessed based on studying the form of the Schwarzschild-de Sitter spacetime (and subsequently shown to be consistent throughout the calculations involving all 38 Newman-Penrose equations), these fall-offs all agree with the corresponding results from assuming a smooth conformal compactifiability \cite{Szabados}.}:
\begin{align}\label{Lambdapositivemass}
M_{\Lambda}:=M_B+\frac{\Lambda}{3A}\int{\left(\oint{K|\sigma^o|^2d^2S}\right)du},
\end{align}
where $M_B=-A^{-1}\oint{(\Psi^o_2+\sigma^o\dot{\bar{\sigma}}^o)d^2S}$ is the Bondi mass in an asymptotically flat universe and $A=4\pi$ is the area of a round unit 2-sphere. In the above expression, $\Psi^o_2$ is the leading order term of the dyad component of the Weyl spinor $\Psi_2$ when expanded in inverse powers of $r$ away from null infinity $\mathcal{I}$, and $\sigma^o$ is the leading order term for the spin coefficient $\sigma$. The symbol $K$ is the Gauss curvature for the topological 2-spheres of constant $u$ on $\mathcal{I}$. With $\Lambda\neq0$, we would have $\mathcal{I}$ being non-conformally flat when $\sigma^o\neq0$, and consequently $K$ is not 1 but is instead 1 plus terms involving $\sigma^o$ with an overall factor of $\Lambda$ [Eq. (\ref{oneplussomething})] \footnote{In solving the 38 Newman-Penrose equations in Ref. \cite{Vee2016}, the expression for $K$ arises from the spin coefficient equation involving $\delta\alpha$ and $\delta'\alpha'$. In general, $K$ equals to $2\Theta(\theta,\phi)$ plus those new terms involving $\Lambda$ and $\sigma^o$ in Eq. (\ref{oneplussomething}). Here, $\Theta(\theta,\phi)$ is a free function of the two angular coordinates intrinsic to the 2-surfaces of constant $u$ on $\mathcal{I}$, which is introduced as a ``constant of integration''. As was stipulated in Ref. \cite{Vee2016}, the choice of setting $\Theta(\theta,\phi)$ to 1/2 implies that $K=1$ so that this is associated with a round unit 2-sphere, if $\Lambda$ or $\sigma^o$ is zero. Hence, the presence of $\sigma^o$ \emph{and} $\Lambda$ would induce warpings on the round unit 2-sphere, resulting in a non-conformally flat $\mathcal{I}$ as one finds that the Cotton tensor of $\mathcal{I}$ becomes non-zero.}. The correction term to the Bondi mass $M_B$ from the asymptotically flat case to get $M_\Lambda$ for the $\Lambda>0$ case was proposed as in Eq. (\ref{Lambdapositivemass}) in order to guarantee that the resulting mass-loss formula:
\begin{eqnarray}
\frac{dM_{\Lambda}}{du}
&=&-\frac{1}{A}\oint{\bigg(|\dot{\sigma}^o|^2+\frac{\Lambda}{3}|\eth'\sigma^o|^2+\frac{2\Lambda^2}{9}|\sigma^o|^4}+\frac{\Lambda^2}{18}\textrm{Re}(\bar{\sigma}^o\Psi^o_0)\bigg)d^2S\label{0Bondimasslosspsi0}
\end{eqnarray}
is manifestly positive-definite with an overall negative sign when $\Psi^o_0$ (the leading order term of $\Psi_0$ which represents incoming radiation from elsewhere) is zero, so that the gravitational waves emitted by the isolated system carry away a positive-definite energy and reduce the mass $M_\Lambda$. This mass-loss formula arises from the same Bianchi identity where Newman and Unti obtained for the asymptotically flat case, i.e. the one involving $D'\Psi_2$ --- which incidentally, corresponds to the ``supplementary condition'' upon which Bondi got the mass-loss formula \cite{Bondi62}.

Around the same time when this work in Ref. \cite{Vee2016} was being carried out, a Bondi-type mass with $\Lambda$ was independently given by Chru{\'s}ciel and Ifsits \cite{Chrusciel}, containing a correction term which they referred to as a ``renormalised volume of the null hypersurface'' (with alternative ways of re-expressing and re-defining terms, as well as possibly moving the terms across the ``$=$'' sign, in their ``balance formula''). Although we presently do not know how to directly relate these two proposals of the Bondi mass with $\Lambda>0$ in Refs. \cite{Vee2016} and \cite{Chrusciel} (as the notations and conventions used in these papers are disparate), it seemed plausible (and natural) to have a generalisation of $M_B$ to the $\Lambda>0$ case with new terms involving $\Lambda$ \footnote{A need for corrections to the mass $M_\Lambda$ is particularly compelling for the $\Lambda<0$ case, where the term involving $\Lambda|\eth'\sigma^o|^2$ in Eq.~(\ref{0Bondimasslosspsi0}) is \emph{negative-definite}.}. The correction term in Eq. (\ref{Lambdapositivemass}) proposed in Ref. \cite{Vee2016} may also be interpreted as a volume integral since it involves an integral over a 2-surface of constant $u$ on $\mathcal{I}$ with a further integration over $u$ on $\mathcal{I}$. A research direction would then be to reconcile these two approaches and validate the connection between these two ``volume correction terms''.

However, the nature of an integral over $u$ (in an expression describing the mass of an isolated system at some particular moment $u=u_0$) implies that one would allegedly require information of the system not just at that moment, but also of its past history from say $u=0$, upon where that integral is carried out from --- unless one is able to express that volume integral entirely in terms of local expressions at $u=u_0$. The presence of such a volume integral would be perceived as a significant departure from the asymptotically flat case. Unfortunately, one faces formidable (but hopefully not irreconcilable) difficulty in attempting to pin down an exact notion of the Bondi mass when $\Lambda\neq0$, because this $\Lambda$ destroys the asymptotic symmetries which formed the concrete basis for a Hamiltonian treatment of the Bondi mass when $\Lambda=0$ \cite{Vee2017d}. A ramification of this is the ambiguity in definitively defining the Bondi mass, as demonstrated by the undesirable freedom for moving terms around in Eq. (\ref{0Bondimasslosspsi0}) to be absorbed into the definition of $M_\Lambda$, or even adding conceivably arbitrary terms on both sides.

Ergo, the goal of this paper is to provide a way of not only obtaining the mass-loss formula, but also argue for a reasonable generalisation of the Bondi mass with $\Lambda>0$ \footnote{We will find that the $\Lambda<0$ case would not give a manifestly positive-definite expression for the energy carried away by gravitational waves, because the terms with a factor of $\Lambda$ would then have the opposite sign.}. We will do this by applying an integral formula on a hypersurface that was found by Frauendiener, using the Nester-Witten identity \cite{Fra97}. The idea would be to apply the asymptotic expansions for the spin coefficients with $\Lambda\in\R$ that were recently worked out in Ref. \cite{Vee2016} and plug them into the integral formula. This was one of the applications illustrated by Frauendiener, viz. to produce the Bondi mass-loss formula for the asymptotically flat case. The Nester-Witten identity may be expressed in the form ``$dL=S+E$'', and it is this geometric object (a 3-form) ``$dL$'' which involves a differential of ``$L$'' (a 2-form) that becomes the expression for the rate of change of the Bondi mass $M_B$. We adopt the exact steps and find that with $\Lambda\in\R$, this geometric object ``$dL$'' turns out to be \emph{precisely the same expression, $dM_B/du$}. Incidentally, we find that the integral formula yields the identical mass-loss formula with $\Lambda\in\R$ that was first reported in Ref. \cite{Vee2016}, as anticipated.

In the next section, we present the pair of integral formulae on a hypersurface from Ref. \cite{Fra97} and express them in a form that would be true for any $\Lambda\in\R$. With this, we can derive the expected mass-loss formula with $\Lambda\in\R$ in Section \ref{main}. To do so, a number of technical steps are required, which are organised into four subsections. With a generalisation of the Bondi mass to include $\Lambda\in\R$ being just $M_\Lambda=M_B$ (as this is the expression appearing in the geometric object ``$dL$'' for any $\Lambda\in\R$), the sole term that is not \emph{manifestly} positive-definite (in the absence of incoming radiation from elsewhere) for the $\Lambda>0$ case involves the Gauss curvature $K$ of the topological 2-spheres of constant $u$ on $\mathcal{I}$. We show in Section \ref{Gaussidentity} that we can express this in terms of $\eth'\sigma^o$ and $\eth\sigma^o$, using an identity for the commutator of the $\eth$ and $\eth'$ operators \cite{Pen87}. With this, we see that if $\sigma^o$ (which is a spin-weighted quantity with spin-weight $s=2$ \cite{Pen87,Don}) only has $l=2$ components (which represent the quadrupole terms in a multipole expansion of the compact isolated source), then the mass-loss for $M_\Lambda=M_B$ in the $\Lambda>0$ case is \emph{manifestly} positive-definite, viz. \emph{quadrupole gravitational waves carry away positive-definite energy}. Several inequalities are given, which would guarantee this manifestly positive-definite property for a general $\sigma^o$, in a universe with $\Lambda>0$.

To recapitulate on the behaviour of empty asymptotically de Sitter spacetimes, we briefly summarise the main points from Ref. \cite{Vee2016} in Appendix \ref{recapi}. There, we state the assumptions used to derive these asymptotic solutions, as well as what free functions are left in order to specify a unique solution. The stipulations on the fall-offs for the spin coefficients and unknown functions in the null tetrad used in Ref. \cite{Vee2016}, together with the vanishing of the Weyl spinor on $\mathcal{I}$ [i.e. without having to additionally assume that $\Psi_0=\Psi^o_0r^{-5}+O(r^{-6})$ \cite{Vee2017c}] are consistent with assuming a smooth conformal compactifiablity: this can be clearly seen by an application of a boost transformation of the $\vec{l}$ and $\vec{n}$ null vectors (this boost transformation was discussed explicitly in Ref. \cite{Vee2017}, where there asymptotic solutions were expressed in a Szabados-Tod-type null tetrad), together with a null rotation about $\vec{l}$ to arrive at the results reported by Szabados and Tod \cite{Szabados}. (This null rotation gets rid of $\omega$ so that $\vec{m}$ and $\vec{\bar{m}}$ do not have a $\vec{\partial}_r$ component. Incidentally, such a null rotation is carried out in Section \ref{nullrot} because the integral formulae by Frauendiener assumed such $\vec{m}$ and $\vec{\bar{m}}$.) Lastly, we give in Appendix \ref{appendix}: a proof that the area element of the topological 2-sphere of constant $u$ on $\mathcal{I}$, denoted by $d^2S$, is independent of $u$. This implies that the term involving $\partial(d^2S)/\partial u$ in the proposal for $M_\Lambda$ in Ref. \cite{Vee2016} [see Eq. (126) in that paper] is zero.

\section{The integral formulae on a hypersurface, arising from the Nester-Witten identity}\label{preliminary}

Frauendiener derived a pair of integral formulae on some hypersurface $\Sigma$ based on the Nester-Witten identity of the form ``$dL=S+E$'' \cite{Fra97} \footnote{Each of these versions of the integral formulae can be obtained from the other by the so-called ``prime operation'' of the spinor dyads \cite{Don}. Essentially, one switches between the outgoing null direction to the incoming one, and vice versa.}:
\begin{align}
\frac{d}{ds}\oint{\rho\phi d^2A}&=\oint{\rho\frac{d{\phi}}{ds}d^2A}+\oint{Z\phi(\sigma\bar{\sigma}-\rho^2)d^2A}+\oint{Z'\phi\left(\frac{\mathcal{R}}{4}-\tau\bar{\tau}\right)d^2A}\nonumber\\&\phantom{=}-\frac{1}{2}\oint{\phi l^aG_a^{\ b}p_{bc}Z^cd^2A},\textrm{ and}\label{int1}\\
\frac{d}{ds}\oint{\rho'\phi'd^2A}&=\oint{\rho'\frac{d{\phi}'}{ds}d^2A}+\oint{Z'\phi'(\sigma'\bar{\sigma}'-\rho'^2)d^2A}+\oint{Z\phi'\left(\frac{\mathcal{R}}{4}-\tau'\bar{\tau}'\right)d^2A}\nonumber\\&\phantom{=}+\frac{1}{2}\oint{\phi'n^aG_a^{\ b}p_{bc}Z^cd^2A}.\label{int2}
\end{align}
The left-hand sides are ``$dL$'', with ``$E$'' being the term involving the Einstein tensor $G_a^{\ b}$ on the right-hand sides, and the rest of the terms on the right-hand sides are ``$S$''. The Greek letters $\rho,\rho',\sigma,\sigma',\tau,\tau'$ that appear in these formulae are the usual spin coefficients of the Newman-Penrose formalism. The notations and conventions used by Frauendiener are (probably almost) consistent with those in Penrose-Rindler \cite{Pen87,Pen88} \footnote{Private communication with J{\"o}rg Frauendiener.} for the case with zero cosmological constant, and with Refs. \cite{Vee2016,Vee2017} for the case with a cosmological constant $\Lambda\in\R$ (apart from the null tetrad, but this can be reconciled via a null rotation around $\vec{l}$ --- see Section \ref{nullrot}).

In the setup by Frauendiener, the 3-d hypersurface $\Sigma$ under consideration is assumed to be foliated by a set of topological 2-spheres $S_s$ (each value of $s$ labels one such surface), with $\mathcal{R}$ being the scalar curvature of $S_s$ \footnote{Note that our $\mathcal{R}$ here is twice that in Ref. \cite{Fra97}. This is so that our $\mathcal{R}$ would be twice the Gauss curvature. In Ref. \cite{Fra97}, the $\mathcal{R}$ there is four times the Gauss curvature.}. The surface integral is carried out over $S_s$, for some fixed $s$, and its area element is $d^2A=m_{ab}$. An outgoing unit normal to these topological 2-spheres is denoted by $\hat{Z}$ (but one can certainly consider any arbitrary length $\vec{Z}$ parallel to $\hat{Z}$ --- in particular if $\vec{Z}$ is null, then it always has zero length), which may be decomposed into $\vec{Z}=Z\vec{l}+Z'\vec{n}
$, and the $s$-derivative is a directional derivative along $\vec{Z}$. Well, we have here $\{\vec{l}$, $\vec{n}$, $\vec{m}$, $\vec{\bar{m}}\}$ forming a  null tetrad (the first two are orthogonal to $S_s$ whilst the latter two are tangent to $S_s$), with the induced area elements given by $p_{ab}=l_an_b-n_al_b$ and $m_{ab}=-i(m_a\bar{m}_b-\bar{m}_am_b)$. The functions $\phi$ and $\phi'$ are type $(-1,-1)$ and $(1,1)$ functions \cite{Pen87} on $\Sigma$. Whilst these functions are left in there in Ref. \cite{Fra97}, they should just correspond to a boost transformation of the $\vec{l}$ and $\vec{n}$ null vectors. In other words, one can just set $\phi$ and $\phi'$ to 1 (but we will for the moment, leave these $\phi$ and $\phi'$ in there).

To deal with the case with a cosmological constant $\Lambda\in\R$, we need a few things here and re-express these integral formulae. Firstly, Frauendiener assumed that the spin coefficients $\gamma$ and $\gamma'$ vanish on $\Sigma$, as he imposed that the spinor dyads $o^A$ and $\iota^A$ obey the propagations $Do^A=0$ and $D'\iota^A=0$. Such conditions however, cannot be satisfied when there is $\Lambda\in\R$ (in particular, it was found in Ref. \cite{Vee2016} that $\gamma=O(r)$ due to $\Lambda$), and so these terms involving $\gamma$ and $\gamma'$ must be restored into the integral formulae, viz.
\begin{align}
&\oint{2\rho\phi(Z'\textrm{Re}(\gamma)-Z\textrm{Re}(\gamma'))d^2A}\textrm{ into Eq. (\ref{int1}), and}\\
&\oint{2\rho'\phi'(Z\textrm{Re}(\gamma')-Z'\textrm{Re}(\gamma))d^2A}\textrm{ into Eq. (\ref{int2})}.
\end{align}
Apart from that, the Einstein tensor is not zero but is given by the Einstein field equations as $G_a^{\ b}=-\Lambda\delta_a^{\ b}$. With this as well as $p_{bc}=l_bn_c-n_bl_c$, $Z^c=Zl^c+Z'n^c$, and the orthonormalisation of $\vec{l}$ and $\vec{n}$ ($l^an_a=1$, $l^al_a=n^an_a=0$), the last term in Eq. (\ref{int1}) becomes:
\begin{align}
-\frac{1}{2}\oint{\phi l^aG_a^{\ b}p_{bc}Z^cd^2A}&=\frac{1}{2}\oint{\Lambda\phi l^a\delta_a^{\ b}(l_bn_c-n_bl_c)(Zl^c+Z'n^c)d^2A}\\
&=\frac{1}{2}\oint{\Lambda\phi l^b(Zl_b-Z'n_b)d^2A}\\
&=-\frac{1}{2}\oint{Z'\Lambda\phi d^2A}.
\end{align}
Similarly, the last term in Eq. (\ref{int2}) becomes:
\begin{align}
\frac{1}{2}\oint{\phi'n^aG_a^{\ b}p_{bc}Z^cd^2A}&=-\frac{1}{2}\oint{Z\Lambda\phi'd^2A}.
\end{align}
With these to account for $\Lambda\in\R$ (as far as we are able to detect), the pair of integral formulae from Eqs. (\ref{int1}) and (\ref{int2}) are:
\begin{align}
\frac{d}{ds}\oint{\rho\phi d^2A}&=\oint{\rho\frac{d{\phi}}{ds}d^2A}+\oint{\rho\phi(2Z'\textrm{Re}(\gamma)-Z(2\textrm{Re}(\gamma')+\rho))d^2A}+\oint{Z\phi\sigma\bar{\sigma}d^2A}\nonumber\\&\phantom{=}+\oint{Z'\phi\left(\frac{\mathcal{R}}{4}-\frac{\Lambda}{2}-\tau\bar{\tau}\right)d^2A},\textrm{ and}\label{int11}\\
\frac{d}{ds}\oint{\rho'\phi'd^2A}&=\oint{\rho'\frac{d{\phi}'}{ds}d^2A}+\oint{\rho'\phi'(2Z\textrm{Re}(\gamma')-Z'(2\textrm{Re}(\gamma)+\rho'))d^2A}+\oint{Z'\phi'\sigma'\bar{\sigma}'d^2A}\nonumber\\&\phantom{=}+\oint{Z\phi'\left(\frac{\mathcal{R}}{4}-\frac{\Lambda}{2}-\tau'\bar{\tau}'\right)d^2A}.\label{int22}
\end{align}

Incidentally, the left-hand side of Eq. (\ref{int11}) (which is the ``$dL$'' of the Nester-Witten identity ``$dL=S+E$'') can be expanded as
\begin{align}
\frac{d}{ds}\oint{\rho\phi d^2A}=\oint{\frac{d}{ds}(\rho\phi d^2A)}=\oint{\left(\frac{d\rho}{ds}\phi d^2A+\rho\phi\frac{d(d^2A)}{ds}+\rho\frac{d\phi}{ds}d^2A\right)},\label{dphids}
\end{align}
so we find that the term involving the $s$-derivative of $\phi$ exactly cancels out the corresponding term on the right-hand side. The remaining terms all involve a factor of $\phi$, and since $\phi$ is an arbitrary function, we must have the following integral formulae [where similar remarks hold for Eq. (\ref{int22})] independent of $\phi$ (and $\phi'$) --- or just set $\phi$ and $\phi'$ to 1 \footnote{The $s$-derivative of the area element $d^2A$ can be evaluated once a choice of $\vec{Z}$ is made, to end up with an expression analogous to Eq. (\ref{areaelement}) below in terms of $Z$ and $Z'$. In that case, since $\phi$ and $\phi'$ are arbitrary, we see that we get  equations for the integrands themselves without having to carry out the integration on the 2-surface $S_s$ for some fixed $s$. We should expect that the resulting differential equations are equivalent to (some of) the Bianchi identities --- but one should carefully revisit the paper in Ref. \cite{Fra97} to go through the entire derivation right from the start. As the purpose of this paper is to study the Bondi mass, we do not go into that discussion here, but instead carry on with the integrated formulae.}:
\begin{align}
\oint{\frac{d\rho}{ds}d^2A}+\oint{\rho\frac{d(d^2A)}{ds}}&=\oint{\rho(2Z'\textrm{Re}(\gamma)-Z(2\textrm{Re}(\gamma')+\rho))d^2A}+\oint{Z\sigma\bar{\sigma}d^2A}\nonumber\\&\phantom{=}+\oint{Z'\left(\frac{\mathcal{R}}{4}-\frac{\Lambda}{2}-\tau\bar{\tau}\right)d^2A},\textrm{ and}\label{int111}\\
\oint{\frac{d\rho'}{ds}d^2A}+\oint{\rho'\frac{d(d^2A)}{ds}}&=\oint{\rho'(2Z\textrm{Re}(\gamma')-Z'(2\textrm{Re}(\gamma)+\rho'))d^2A}+\oint{Z'\sigma'\bar{\sigma}'d^2A}\nonumber\\&\phantom{=}+\oint{Z\left(\frac{\mathcal{R}}{4}-\frac{\Lambda}{2}-\tau'\bar{\tau}'\right)d^2A}.\label{int222}
\end{align}

These are thus, the general pair of integral formulae on a hypersurface based on the Nester-Witten identity, including a cosmological constant $\Lambda\in\R$. With regards to the mass-loss formula, since the mass aspect (at least, for the $\Lambda=0$ case) involving $\Psi^o_2+\sigma^o\dot{\bar{\sigma}}^o$ appears in the spin coefficient $\rho'$, we are most interested in the rate of change of this quantity over different surfaces of constant $u$ on null infinity $\mathcal{I}$. This is the goal of Section \ref{main}, i.e. to evaluate the second integral formula Eq. (\ref{int222}) on $\mathcal{I}$.

As mentioned at the beginning of the section, the left-hand sides of the integral formulae are the ``$dL$'' part of the Nester-Witten identity, ``$dL=S+E$''. The ``$E$'' refers to the Einstein tensor, which in this case is just the term involving $\Lambda$, and the rest is ``$S$''. We will be interested to isolate ``$dL$'' which will be adopted as a generalisation of the Bondi mass $M_\Lambda$ for any $\Lambda\in\R$.

\section{Evaluating the integral formula on null infinity with $\Lambda\in\R$}\label{main}

\subsection{The null tetrad, spin coefficients, and null rotation}\label{nullrot}

The asymptotic solutions with $\Lambda\in\R$ have recently been worked out in Refs. \cite{Vee2016,Vee2017}. The coordinates used there are $u, r, x^\mu$, with $\mu$ labelling the two angular coordinates on $S_s$, $u$ being a retarded null coordinate (a dot above a quantity represents a partial derivative of that quantity with respect to $u$), and $r$ is an affine parameter of the null geodesics generating the outgoing null hypersurfaces $u=$ constant. Null infinity $\mathcal{I}$ is the hypersurface obtained by taking the limit where $r\rightarrow\infty$. Anyway, the null tetrad vectors are \footnote{The null tetrad vectors and spin coefficients taken directly from Refs. \cite{Vee2016,Vee2017} are denoted with a subscript ``Saw''. Then, the ones converted via a null rotation around $\vec{l}$ to be used for evaluating Eq. (\ref{int222}) will have no subscript tagged to them. Incidentally, the Einstein summation convention is assumed for $\mu$, over the two angular coordinates intrinsic to the topological 2-spheres $S_s$.}
\begin{align}
\vec{l}_{\textrm{Saw}}&=\vec{\partial}_r\\
\vec{n}_{\textrm{Saw}}&=\vec{\partial}_u+U\vec{\partial}_r+X^\mu\vec{\partial}_\mu\\
\vec{m}_{\textrm{Saw}}&=\omega\vec{\partial}_r+\xi^\mu\vec{\partial}_\mu\\
\vec{m}_{\textrm{Saw}}&=\bar{\omega}\vec{\partial}_r+\overline{\xi}^\mu\vec{\partial}_\mu,
\end{align}
where
\begin{align}
U&=\frac{\Lambda}{6}r^2-\left(\frac{1}{2}K+\frac{\Lambda}{2}|\sigma^o|^2\right)-\textrm{Re}(\Psi^o_2)r^{-1}+\left(\frac{1}{3}\textrm{Re}(\eth'\Psi^o_1)-\frac{\Lambda}{18}\textrm{Re}(\bar{\sigma}^o\Psi^o_0)\right)r^{-2}+O(r^{-3})\\
X^\mu&=O(r^{-3})\label{Xmu}\\
\omega&=\eth'\sigma^or^{-1}-\left(\sigma^o\eth\bar{\sigma}^o+\frac{1}{2}\Psi^o_1\right)r^{-2}+O(r^{-3})\\
\xi^\mu&=(\xi^\mu)^or^{-1}-\sigma^o(\overline{\xi^\mu})^or^{-2}+O(r^{-3}),
\end{align}
with $K$ being the Gauss curvature of the 2-surfaces of constant $u$ on $\mathcal{I}$:
\begin{align}
\frac{\partial K}{\partial u}&=\frac{2\Lambda}{3}\textrm{Re}(\eth^2\bar{\sigma}^o),\textrm{ i.e.}\\
K&=\Theta(x^\mu)+\frac{2\Lambda}{3}\int{\textrm{Re}(\eth^2\bar{\sigma}^o)du}.\label{oneplussomething}
\end{align}
The free function $\Theta(x^\mu)$ was set to 1 in Ref. \cite{Vee2016} so that when $\Lambda=0$, one then recovers the asymptotically flat result where a conformal transformation of 2-surfaces would turn these surfaces of constant $u$ on $\mathcal{I}$ into \emph{round} 2-spheres \footnote{Later in Section \ref{Gaussidentity} where we derive an identity involving $K$ [that appears in the mass-loss formula Eq. (\ref{masslossformulalah})] in terms of only the $\eth$ and $\eth'$ of $\sigma^o$ [see Eq. (\ref{Gaussidentityequation})], we see that $\Theta(x^\mu)$ plays \emph{no role and does not affect the mass-loss formula}. Therefore, $\Theta(x^\mu)$ does not carry any physical significance with regards to the mass-loss formula, and so we set $\Theta(x^\mu)$ = 1.}. With a non-zero $\Lambda\in\R$ however, these 2-surfaces of constant $u$ on $\mathcal{I}$ are not round 2-spheres. Instead, they are topological 2-spheres. Hence, the $\eth$ and $\eth'$ operators are defined on these topological 2-spheres of constant $u$ on $\mathcal{I}$ \cite{Vee2016,Vee2017}
\begin{align}
\eth\eta:&=(\xi^\mu)^o\frac{\partial\eta}{\partial x^\mu}+2s\bar{\alpha}^o\eta\label{eth}\\
\eth'\eta:&=(\overline{\xi^\mu})^o\frac{\partial\eta}{\partial x^\mu}-2s\alpha^o\eta\label{ethprime},
\end{align}
where $\eta$ is a spin-weighted quantity with spin-weight $s$ \cite{Pen87}, and $\alpha^o$ is the leading order term of the spin coefficient $\alpha$ (when expanded in inverse powers of $r$ away from $\mathcal{I}$) \cite{Vee2016}. For brevity, we sometimes denote the partial derivative by $\partial$ with a subscript indicative of the coordinate, i.e. $\partial/\partial u=\partial_u$, $\partial/\partial r=\partial_r$, $\partial/\partial x^\mu=\partial_\mu$.

Now, the setup by Frauendiener employed a null tetrad where $\vec{m}$ and $\vec{\bar{m}}$ are tangential to $S_s$, i.e. they do not have any $\vec{\partial}_r$ component \cite{Fra97}. In order to convert the null tetrad and spin coefficients found in Refs. \cite{Vee2016,Vee2017} so that we may apply them here to evaluate Eq. (\ref{int222}), we perform a null rotation around $\vec{l}$:
\begin{align}
\vec{l}&=\vec{l}_{\textrm{Saw}}\\
\vec{n}&=\vec{n}_{\textrm{Saw}}+\bar{c}\vec{m}_{\textrm{Saw}}+c\vec{\bar{m}}_{\textrm{Saw}}+c\bar{c}\vec{l}_{\textrm{Saw}}\\
\vec{m}&=\vec{m}_{\textrm{Saw}}+c\vec{l}_{\textrm{Saw}}\\
\vec{\bar{m}}&=\vec{\bar{m}}_{\textrm{Saw}}+\bar{c}\vec{l}_{\textrm{Saw}},
\end{align}
where $c$ is a complex function. Letting $c=-\omega$, the new null tetrad vectors are:
\begin{align}
\vec{l}&=\vec{\partial}_r\label{nrnt1}\\
\vec{n}&=\vec{\partial}_u+(U-\omega\bar{\omega})\vec{\partial}_r+(X^\mu-2\textrm{Re}(\bar{\omega}\xi^\mu))\vec{\partial}_\mu\\
\vec{m}&=\xi^\mu\vec{\partial}_\mu\\
\vec{m}&=\overline{\xi}^\mu\vec{\partial}_\mu.\label{nrnt2}
\end{align}

The relevant spin coefficients (for evaluating the integral formula here) taken from Ref. \cite{Vee2016} are:
\begin{align}
\rho_{\textrm{Saw}}&=-r^{-1}-|\sigma^o|^2r^{-3}+\left(\frac{1}{3}\textrm{Re}(\bar{\sigma}^o\Psi^o_0)-|\sigma^o|^4\right)r^{-5}+O(r^{-6})\\
\rho'_{\textrm{Saw}}&=-\frac{\Lambda}{6}r+\left(\frac{1}{2}K+\frac{\Lambda}{3}|\sigma^o|^2\right)r^{-1}+(\Psi^o_2+\sigma^o\dot{\bar{\sigma}}^o)r^{-2}\nonumber\\&\phantom{=}+\left(\frac{1}{2}K|\sigma^o|^2+\frac{\Lambda}{3}|\sigma^o|^4+\frac{\Lambda}{9}\textrm{Re}\left(\bar{\sigma}^o\Psi^o_0\right)-\frac{1}{2}\eth'\Psi^o_1\right)r^{-3}+O(r^{-4})\\
\sigma_{\textrm{Saw}}&=\sigma^or^{-2}+\left(\sigma^o|\sigma^o|^2-\frac{1}{2}\Psi^o_0\right)r^{-4}+O(r^{-5})\\
\sigma'_{\textrm{Saw}}&=-\frac{\Lambda}{6}\bar{\sigma}^o-\dot{\bar{\sigma}}^or^{-1}-\left(\frac{1}{2}K\bar{\sigma}^o+\frac{\Lambda}{3}\bar{\sigma}|\sigma^o|^2+\frac{\Lambda}{12}\bar{\Psi}^o_0\right)r^{-2}+O(r^{-3})\\
\textrm{Re}(\gamma_{\textrm{Saw}})&=-\frac{\Lambda}{6}r-\frac{1}{2}\textrm{Re}(\Psi^o_2)r^{-2}+\left(\frac{1}{3}\textrm{Re}(\eth'\Psi^o_1)-\frac{\Lambda}{18}\textrm{Re}(\bar{\sigma}^o\Psi^o_0)\right)r^{-3}+O(r^{-4})\\
\gamma'_{\textrm{Saw}}&=0\\
\tau_{\textrm{Saw}}&=O(r^{-3})\\
\tau'_{\textrm{Saw}}&=0\\
\alpha_{\textrm{Saw}}&=\alpha^or^{-1}+\bar{\alpha}^o\bar{\sigma}^or^{-2}+O(r^{-3})\\
\alpha'_{\textrm{Saw}}&=\bar{\alpha}^or^{-1}+\alpha^o\sigma^or^{-2}+O(r^{-3}).
\end{align}
Under the required null rotation, the relevant spin coefficients transform as:
\begin{align}
\rho&=\rho_{\textrm{Saw}}\\
\rho'&=\rho'_{\textrm{Saw}}-\bar{\omega}^2\sigma_{\textrm{Saw}}-2\bar{\omega}\alpha'_{\textrm{Saw}}+\delta\bar{\omega}-\omega D\bar{\omega}\\
\sigma&=\sigma_{\textrm{Saw}}\\
\sigma'&=\sigma'_{\textrm{Saw}}+2\bar{\omega}\alpha_{\textrm{Saw}}-\bar{\omega}^2\rho_{\textrm{Saw}}+\delta'\bar{\omega}-\bar{\omega}D\bar{\omega}\\
\gamma&=\gamma_{\textrm{Saw}}-\omega\alpha_{\textrm{Saw}}+\bar{\omega}\alpha'_{\textrm{Saw}}-\bar{\omega}\tau_{\textrm{Saw}}+\omega\bar{\omega}\rho_{\textrm{Saw}}+\bar{\omega}^2\sigma_{\textrm{Saw}}\\
\gamma'&=0\\
\tau&=\tau_{\textrm{Saw}}-\omega\rho_{\textrm{Saw}}-\bar{\omega}\sigma_{\textrm{Saw}}\\
\tau'&=D\bar{\omega}.
\end{align}
In the above, $D,D',\delta,\delta'$ are directional derivatives along $\vec{l}_{\textrm{Saw}},\vec{n}_{\textrm{Saw}},\vec{m}_{\textrm{Saw}},\vec{\bar{m}}_{\textrm{Saw}}$, respectively. Therefore, the null rotated spin coefficients expressed in terms of the null tetrad given by Eqs. (\ref{nrnt1})-(\ref{nrnt2}) are:
\begin{align}
\rho&=-r^{-1}-|\sigma^o|^2r^{-3}+\left(\frac{1}{3}\textrm{Re}(\bar{\sigma}^o\Psi^o_0)-|\sigma^o|^4\right)r^{-5}+O(r^{-6})\label{lhohwithoutprime}\\
\rho'&=-\frac{\Lambda}{6}r+\left(\frac{1}{2}K+\frac{\Lambda}{3}|\sigma^o|^2\right)r^{-1}+(\Psi^o_2+\sigma^o\dot{\bar{\sigma}}^o+\eth^2\bar{\sigma}^o)r^{-2}\nonumber\\&\phantom{=}+\left(\frac{1}{2}K|\sigma^o|^2+\frac{\Lambda}{3}|\sigma^o|^4+\frac{\Lambda}{9}\textrm{Re}\left(\bar{\sigma}^o\Psi^o_0\right)-\textrm{Re}(\eth'\Psi^o_1)-|\eth'\sigma^o|^2-2\textrm{Re}(\sigma^o\eth'\eth\bar{\sigma}^o)\right)r^{-3}\nonumber\\&\phantom{=}+O(r^{-4})\label{lhoh}\\
\sigma&=\sigma^or^{-2}+\left(\sigma^o|\sigma^o|^2-\frac{1}{2}\Psi^o_0\right)r^{-4}+O(r^{-5})\\
\sigma'&=-\frac{\Lambda}{6}\bar{\sigma}^o-\dot{\bar{\sigma}}^or^{-1}-\left(\frac{1}{2}K\bar{\sigma}^o+\frac{\Lambda}{3}\bar{\sigma}^o|\sigma^o|^2-\eth'\eth\bar{\sigma}^o+\frac{\Lambda}{12}\bar{\Psi}^o_0\right)r^{-2}+O(r^{-3})\\
\textrm{Re}(\gamma)&=-\frac{\Lambda}{6}r-\frac{1}{2}\textrm{Re}(\Psi^o_2)r^{-2}+\left(\frac{1}{3}\textrm{Re}(\eth'\Psi^o_1)-|\eth'\sigma^o|^2-\frac{\Lambda}{18}\textrm{Re}(\bar{\sigma}^o\Psi^o_0)\right)r^{-3}+O(r^{-4})\\
\gamma'&=0\\
\tau&=\eth'\sigma^or^{-2}+O(r^{-3})\\
\tau'&=-\eth\bar{\sigma}^or^{-2}+O(r^{-3}).
\end{align}
Note that in this null tetrad employed by Frauendiener, $\rho'$ is real \cite{Fra97}, so $\Psi^o_2+\sigma^o\dot{\bar{\sigma}}^o+\eth^2\bar{\sigma}^o=\textrm{Re}(\Psi^o_2+\sigma^o\dot{\bar{\sigma}}^o+\eth^2\bar{\sigma}^o)$.

\subsection{The hypersurface with $\vec{Z}=\vec{n}-(U-\omega\bar{\omega})\vec{l}$ and the $s$-derivative of the area element}\label{Zzz}

To obtain the mass-loss formula, the goal is to have a derivative of the mass aspect $\Psi^o_2+\sigma^o\dot{\bar{\sigma}}^o$ [which appears in the $r^{-2}$ order of $\rho'$, see Eq. (\ref{lhoh})] involving different values of $u$ on $\mathcal{I}$. The hypersurface $\mathcal{I}$ is reached by taking the limit $r\rightarrow\infty$, but before taking this limit, we have to work with some hypersurface $\Sigma_r$ having finite $r$ (i.e. $\Sigma_\infty$ is $\mathcal{I}$). The need for a $u$-derivative implies that this should involve a derivative along (the null rotated) $\vec{n}$, i.e. $D'=\nabla_an^a$ (where this $D'$ is along the null rotated $\vec{n}$, not $\vec{n}_{\textrm{Saw}}$). The simplest way to have $D'$ would be to consider a hypersurface $\Sigma_r$ with the vector $\vec{Z}$ (representing the outward or inward flow of the topological 2-spheres $S_s$ that foliate $\Sigma_r$) being $\vec{Z}=\vec{n}=\vec{\partial}_u+(U-\omega\bar{\omega})\vec{\partial}_r+(X^\mu-2\textrm{Re}(\bar{\omega}\xi^\mu))\vec{\partial}_\mu$. Then, since $\vec{Z}=Z\vec{l}+Z'\vec{n}$, we get $Z=0$ and $Z'=1$. With this, the $s$-derivative would be $\partial/\partial u+(U-\omega\bar{\omega})\partial/\partial r+(X^\mu-2\textrm{Re}(\bar{\omega}\xi^\mu))\partial/\partial x_\mu$. This would indeed work (as we have verified), and perhaps requires the least amount of calculations since $Z=0$ would kill off terms involving the Ricci scalar $\mathcal{R}$ and the spin coefficient $\tau'$ in the integral formula as can be seen in Eq. (\ref{int222}).

Now, we wish to follow the setup here with $\Lambda\in\R$ as closely as possible to the asymptotically flat case. With a null $\mathcal{I}$ when $\Lambda=0$, there is then a canonical choice for $\vec{Z}$, viz. $\vec{Z}=\vec{\partial}_u$ as was done by Frauendiener \cite{Fra97}. This choice of $\vec{Z}$ implies that the foliation of the hypersurface under consideration by $S_s$ is the one where $S_s$ are the 2-surfaces of constant $u$ (denoted by $S_u$), whereas other choices of $\vec{Z}$ would foliate the hypersurface differently. As we are interested in the mass-loss formula defined on a constant $u$-cut on $\mathcal{I}$, this is the geometrically relevant choice for $\vec{Z}$.


Ergo, let us consider having $\vec{Z}=\vec{n}-(U-\omega\bar{\omega})\vec{l}=\vec{\partial}_u+(X^\mu-2\textrm{Re}(\bar{\omega}\xi^\mu))\vec{\partial}_\mu$, which is the same $\vec{Z}$ taken by Frauendiener for the asymptotically flat case \cite{Fra97}, i.e. $Z=-U+\omega\bar{\omega}$ and $Z'=1$. As he had explained in his derivation of the integral formulae, the angular terms $\vec{\partial}_\mu$ in $\vec{Z}$ give zero contribution because they are integrated away as boundary terms on the topological 2-sphere (which has no boundary). This $\vec{Z}$ is thus effectively $\vec{\partial}_{u}$. We shall proceed to evaluate the $s$-derivative (which is $Z^a\nabla_a$) of the area element $d^2A$ of the 2-surfaces that foliate the hypersurface $\Sigma_r$. In the asymptotically flat case, the $s$-derivative of such area element is zero if one foliates $\mathcal{I}$ by round spheres. With $\Lambda\neq0$ however, $\mathcal{I}$ \emph{cannot} be foliated by round spheres if $\sigma^o\neq0$, i.e. when there is Bondi news \cite{Vee2016}, and consequently we need to work this out explicitly.

Firstly, this area element for the topological 2-sphere $S_s$ is $d^2A=-im_a\wedge\bar{m}_b$ \footnote{The indices on the right-hand side are inserted to keep track of the one-forms $m_a$ and $\bar{m}_b$, especially in the calculations that follow.}. Since the $s$-derivative is a directional derivative along $\vec{Z}$, this is equivalent to the Lie derivative along $\vec{Z}$:
\begin{align}
\frac{d(d^2A)}{ds}&=-i\pounds_{\vec{Z}}(m_a\wedge\bar{m}_b)\\
&=-i\pounds_{\vec{Z}}m_a\wedge\bar{m}_b-im_a\wedge\pounds_{\vec{Z}}\bar{m}_b.
\end{align}
Well,
\begin{align}
\pounds_{\vec{Z}}m_a&=Z^b\nabla_bm_a+m_b\nabla_aZ^ b\\
&=(n^b-(U-\omega\bar{\omega})l^b)\nabla_bm_a+m_b\nabla_a(n^b-(U-\omega\bar{\omega})l^b)\\
&=D'm_a-(U-\omega\bar{\omega})Dm_a+m_b\nabla_an^b-(U-\omega\bar{\omega})m_b\nabla_al^b,
\end{align}
using $\nabla_a(fV^b)=f\nabla_aV^b+V^b\nabla_af$ and $m_bl^b=0$. With $D'm_a=2i\textrm{Im}(\gamma)m_a$ and $Dm_a=-2i\textrm{Im}(\gamma')m_a$ \cite{Don},
\begin{align}
\pounds_{\vec{Z}}m_a&=2i\textrm{Im}(\gamma)m_a+2i(U-\omega\bar{\omega})\textrm{Im}(\gamma')m_a+m_b\nabla_an^b-(U-\omega\bar{\omega})m_b\nabla_al^b.
\end{align}
Then,
\begin{align}
(\pounds_{\vec{Z}}m_a)\bar{m}^a&=-2i\textrm{Im}(\gamma)-2i(U-\omega\bar{\omega})\textrm{Im}(\gamma')+m_b\delta'n^b-(U-\omega\bar{\omega})m_b\delta'l^b\\
&=-2i\textrm{Im}(\gamma)-2i(U-\omega\bar{\omega})\textrm{Im}(\gamma')+\bar{\rho}'-(U-\omega\bar{\omega})\rho,\textrm{ and}\\
(\pounds_{\vec{Z}}m_a)m^a&=m_b\delta n^b-(U-\omega\bar{\omega})m_b\delta l^b\\
&=\bar{\sigma}'-(U-\omega\bar{\omega})\sigma,\textrm{ i.e.}\\
\pounds_{\vec{Z}}m_a&=(2i\textrm{Im}(\gamma)+2i(U-\omega\bar{\omega})\textrm{Im}(\gamma')-\bar{\rho}'+(U-\omega\bar{\omega})\rho)m_a-(\bar{\sigma}'-(U-\omega\bar{\omega})\sigma)\bar{m}_a,
\end{align}
where the spin coefficients $\rho$, $\rho'$, $\sigma$ and $\sigma'$ are defined as $m_b\delta'l^b$, $\bar{m}_b\delta n^b$, $m_b\delta l^b$ and $\bar{m}_b\delta'n^b$ respectively, with $\vec{m},$ $\vec{\bar{m}}$ satisfying the orthonormalisation $m^a\bar{m}_a=-1$, $m^am_a=\bar{m}^a\bar{m}_a=0$ \cite{Don}. Similarly (or just taking the complex conjugate), we have
\begin{align}
\pounds_{\vec{Z}}\bar{m}_a&=-(\sigma'-(U-\omega\bar{\omega})\bar{\sigma})m_a+(-2i\textrm{Im}(\gamma)-2i(U-\omega\bar{\omega})\textrm{Im}(\gamma')-\rho'+(U-\omega\bar{\omega})\bar{\rho})\bar{m}_a.
\end{align}
Hence,
\begin{align}
\frac{d(d^2A)}{ds}&=(-\bar{\rho}'+(U-\omega\bar{\omega})\rho-\rho'+(U-\omega\bar{\omega})\bar{\rho})(-im_a\wedge\bar{m}_b)\\
&=2\textrm{Re}((U-\omega\bar{\omega})\rho-\rho')d^2A\\
&=2((U-\omega\bar{\omega})\rho-\rho')d^2A,\label{areaelement}
\end{align}
where we recall that $\rho$ and $\rho'$ in Eqs. (\ref{lhohwithoutprime}) and (\ref{lhoh}) respectively are real in this null tetrad given by Eqs. (\ref{nrnt1})-(\ref{nrnt2}).

With this $\vec{Z}=\vec{n}-(U-\omega\bar{\omega})\vec{l}$, the integral formulae in Eqs. (\ref{int111}) and (\ref{int222}) are:
\begin{align}
&\ \oint{\frac{d\rho}{ds}d^2A}+2\oint{\rho((U-\omega\bar{\omega})\rho-\rho')d^2A}=\nonumber\\&\ \oint{\rho(2\textrm{Re}(\gamma)+(U-\omega\bar{\omega})\rho)d^2A}-\oint{(U-\omega\bar{\omega})\sigma\bar{\sigma}d^2A}+\oint{\left(\frac{\mathcal{R}}{4}-\frac{\Lambda}{2}-\tau\bar{\tau}\right)d^2A},\textrm{ and}\label{finalintwithoutprime}\\
&\ \oint{\frac{d\rho'}{ds}d^2A}+2\oint{\rho'((U-\omega\bar{\omega})\rho-\rho')d^2A}=\nonumber\\&\  -\oint{\rho'(2\textrm{Re}(\gamma)+\rho')d^2A}+\oint{\sigma'\bar{\sigma}'d^2A}-\oint{(U-\omega\bar{\omega})\left(\frac{\mathcal{R}}{4}-\frac{\Lambda}{2}-\tau'\bar{\tau}'\right)d^2A}.\label{finalint}
\end{align}

\subsection{Evaluating the first integral formula up to order $r^{-4}$}

The mass-loss formula arises from the second integral formula Eq. (\ref{finalint}) (at the $r^{-2}$ order of the integrands). This choice of $\vec{Z}$ requires knowledge of the integral of $\mathcal{R}$ up to order $r^{-4}$, because the lowest order term of $Z=-U+\omega\bar{\omega}$ is $-\Lambda r^2/6$. Interestingly, this can be obtained from the first integral formula Eq. (\ref{finalintwithoutprime}). By expressing $\mathcal{R}$ in terms of an expansion from $\mathcal{I}$ in inverse powers of $r$,
\begin{align}
\mathcal{R}=2Kr^{-2}+\mathcal{R}^o_3r^{-3}+\mathcal{R}^o_4r^{-4}+O(r^{-5}),
\end{align}
our goal here is to work out the integrals of $\mathcal{R}^o_3$ and $\mathcal{R}^o_4$ using Eq. (\ref{finalintwithoutprime}) \footnote{These higher order terms of $\mathcal{R}$ may be derived using the spin coefficient equation involving the $\delta$ and $\delta'$ derivatives of $\alpha$ and $\alpha'$ \cite{Pen87}. This gives $\mathcal{R}^o_3=4\textrm{Re}(\eth^2\bar{\sigma}^o)$ and $\mathcal{R}^o_4=2K|\sigma^o|^2-4|\eth'\sigma^o|^2-4\textrm{Re}(\sigma^o\eth'\eth\bar{\sigma}^o)$, in agreement with the results found in Eqs. (\ref{R3}) and (\ref{R4}) using the first integral formula.}.

Well, we can calculate each integrand term in Eq. (\ref{finalintwithoutprime}) up to order $O(r^{-4})$, and the results are listed below:
\begin{align}
\frac{d\rho}{ds}&=-2\textrm{Re}(\sigma^o\dot{\bar{\sigma}}^o)r^{-3}+O(r^{-5})\\
2\rho((U-\omega\bar{\omega})\rho-\rho')&=2\textrm{Re}(\sigma^o\dot{\bar{\sigma}}^o+\eth^2\bar{\sigma}^o)r^{-3}\nonumber\\&\phantom{=}-4\left(|\eth'\sigma^o|^2+\textrm{Re}(\sigma^o\eth'\eth\bar{\sigma}^o)+\frac{1}{3}\textrm{Re}(\eth'\Psi^o_1)\right)r^{-4}+O(r^{-5})\\
\rho(2\textrm{Re}(\gamma)+(U-\omega\bar{\omega})\rho)&=\frac{\Lambda}{2}+\left(-\frac{1}{2}K+\frac{\Lambda}{6}|\sigma^o|^2\right)r^{-2}\nonumber\\&\phantom{=}+\left(-K|\sigma^o|^2-\frac{\Lambda}{6}|\sigma^o|^4+|\eth'\sigma^o|^2-\frac{\Lambda}{6}\textrm{Re}(\bar{\sigma}^o\Psi^o_0)-\frac{1}{3}\textrm{Re}(\eth'\Psi^o_1)\right)r^{-4}\nonumber\\&\phantom{=}+O(r^{-5})\\
-(U-\omega\bar{\omega})\sigma\bar{\sigma}&=-\frac{\Lambda}{6}|\sigma^o|^2r^{-2}+\left(\frac{1}{2}K|\sigma^o|^2+\frac{\Lambda}{6}|\sigma^o|^4+\frac{\Lambda}{6}\textrm{Re}(\bar{\sigma}^o\Psi^o_0)\right)r^{-4}+O(r^{-5})\\
\frac{\mathcal{R}}{4}-\frac{\Lambda}{2}-\tau\bar{\tau}&=-\frac{\Lambda}{2}+\frac{1}{2}Kr^{-2}+\frac{1}{4}\mathcal{R}^o_3r^{-3}+\left(\frac{1}{4}\mathcal{R}^o_4-|\eth'\sigma^o|^2\right)r^{-4}+O(r^{-5}).
\end{align}
Therefore, from the first integral formula Eq. (\ref{finalintwithoutprime}), the lowest non-trivial order in powers of $r^{-1}$ is:
\begin{align}\label{R3}
\oint{\mathcal{R}^o_3}d^2A=0.
\end{align}
The next order gives:
\begin{align}\label{R4}
\oint{\mathcal{R}^o_4}d^2A=\oint{2K|\sigma^o|^2}d^2A.
\end{align}

\subsection{Evaluating the second integral formula up to order $r^{-2}$}

Let us now work out all the individual integrand terms in the second integral formula Eq. (\ref{finalint}), up to order $r^{-2}$. Upon evaluating the second integral formula, we will find that all terms of lower orders of $r$ cancel out, leaving a relationship at order $r^{-2}$. Note that $d^2A$ is of order $r^2$, so this will result in an integral formula on $\Sigma_r$ containing terms independent of $r$ plus $O(r^{-1})$ terms. In the limit where $r\rightarrow\infty$, we will get the desired mass-loss formula with $\Lambda\in\R$ where the hypersurface $\Sigma_\infty$ is null infinity $\mathcal{I}$ [and the $O(r^{-1})$ terms all vanish].

Here are the results of the calculations:
\begin{align}
\frac{d\rho'}{ds}&=\frac{\Lambda}{3}\textrm{Re}(2\sigma^o\dot{\bar{\sigma}}^o+\eth^2\bar{\sigma}^o)r^{-1}+\frac{\partial}{\partial u}(\Psi^o_2+\sigma^o\dot{\bar{\sigma}}^o+\eth^2\bar{\sigma}^o)r^{-2}+O(r^{-3})\label{yuiop}\\
2\rho'((U-\omega\bar{\omega})\rho-\rho')&=\frac{\Lambda}{3}\textrm{Re}(\sigma^o\dot{\bar{\sigma}}^o+\eth^2\bar{\sigma}^o)r^{-1}\nonumber\\&\phantom{=}-\frac{2\Lambda}{3}\left(|\eth'\sigma^o|^2+\textrm{Re}(\sigma^o\eth'\eth\bar{\sigma}^o)+\frac{1}{3}\textrm{Re}(\eth'\Psi^o_1)\right)r^{-2}+O(r^{-3})\\
-\rho'(2\textrm{Re}(\gamma)+\rho')&=-\frac{\Lambda^2}{12}r^2+\left(\frac{\Lambda}{3}K+\frac{2\Lambda^2}{9}|\sigma^o|^2\right)\nonumber\\&\phantom{=}+\left(\frac{\Lambda}{2}\textrm{Re}(\Psi^o_2)+\frac{2\Lambda}{3}\textrm{Re}(\sigma^o\dot{\bar{\sigma}}^o+\eth^2\bar{\sigma}^o)\right)r^{-1}\nonumber\\&\phantom{=}+\bigg(-\frac{1}{4}K^2+\frac{\Lambda^2}{9}|\sigma^o|^4-\Lambda|\eth'\sigma^o|^2-\frac{4\Lambda}{3}\textrm{Re}(\sigma^o\eth'\eth\bar{\sigma}^o)+\frac{\Lambda^2}{18}\textrm{Re}(\bar{\sigma}^o\Psi^o_0)\nonumber\\&\phantom{=++}-\frac{5\Lambda}{9}\textrm{Re}(\eth'\Psi^o_1)\bigg)r^{-2}+O(r^{-3})\\
\sigma'\bar{\sigma}'&=\frac{\Lambda^2}{36}|\sigma^o|^2+\frac{\Lambda}{3}\textrm{Re}(\sigma^o\dot{\bar{\sigma}}^o)r^{-1}\nonumber\\&\phantom{=}+\left(|\dot{\sigma}^o|^2+\frac{\Lambda}{6}K|\sigma^o|^2+\frac{\Lambda^2}{9}|\sigma^o|^4-\frac{\Lambda}{3}\textrm{Re}(\sigma^o\eth'\eth\bar{\sigma}^o)+\frac{\Lambda^2}{36}\textrm{Re}(\bar{\sigma}^o\Psi^o_0)\right)r^{-2}\nonumber\\&\phantom{=}+O(r^{-3})\\
-(U-\omega\bar{\omega})\bigg(\frac{\mathcal{R}}{4}-\frac{\Lambda}{2}&-\tau'\bar{\tau}'\bigg)=\frac{\Lambda^2}{12}r^2-\left(\frac{\Lambda}{3}K+\frac{\Lambda^2}{4}|\sigma^o|^2\right)-\frac{\Lambda}{2}\left(\textrm{Re}(\Psi^o_2)+\frac{1}{12}\mathcal{R}^o_3\right)r^{-1}\nonumber\\&\phantom{-\tau'\bar{\tau}'\bigg)=}\bigg(\frac{1}{4}K^2+\frac{\Lambda}{4}K|\sigma^o|^2-\frac{\Lambda}{24}\mathcal{R}^o_4-\frac{\Lambda}{3}|\eth'\sigma^o|^2-\frac{\Lambda^2}{36}\textrm{Re}(\bar{\sigma}^o\Psi^o_0)\nonumber\\&\phantom{-\tau'\bar{\tau}'\bigg)=+}+\frac{\Lambda}{6}\textrm{Re}(\eth'\Psi^o_1)\bigg)r^{-2}+O(r^{-3}).
\end{align}

Incidentally, using the definitions of the $\eth$ and $\eth'$ operators in Eqs. (\ref{eth}) and (\ref{ethprime}), we find that:
\begin{align}
\partial_u\eth^2\bar{\sigma}^o&=\partial_u\left[(\xi^{\mu})^o\partial_\mu(\eth\bar{\sigma}^o)-2\bar{\alpha}^o\eth\bar{\sigma}^o\right]\\
&=(\dot{\xi}^\mu)^o\partial_\mu(\eth\bar{\sigma}^o)-2\dot{\bar{\alpha}}^o\eth\bar{\sigma}^o+(\xi^\mu)^o\partial_\mu(\partial_u(\eth\bar{\sigma}^o))-2\bar{\alpha}^o\partial_u(\eth\bar{\sigma}^o)\\
&=-\frac{\Lambda}{3}\sigma^o\left[(\overline{\xi^\mu})^o\partial_\mu(\eth\bar{\sigma}^o)+2\alpha^o\eth\bar{\sigma}^o\right]-\frac{\Lambda}{3}|\eth'\sigma^o|^2+\eth(\partial_u(\eth\bar{\sigma}^o))\\
&=-\frac{\Lambda}{3}\sigma^o\eth'\eth\bar{\sigma}^o-\frac{\Lambda}{3}|\eth'\sigma^o|^2+\eth(\partial_u(\eth\bar{\sigma}^o)).
\end{align}
In the above, we went from the second to the third line by making use of the $u$-derivatives of $(\xi^\mu)^o$ and $\alpha^o$ being
\begin{align}
(\dot{\xi}^\mu)^o&=-\frac{\Lambda}{3}\sigma^o(\overline{\xi^\mu})^o\\
\dot{\alpha}^o&=\frac{\Lambda}{3}\bar{\alpha}^o\bar{\sigma}^o+\frac{\Lambda}{6}\eth\bar{\sigma}^o,
\end{align}
respectively \cite{Vee2016,Vee2017}. (These follow from the equations of the Newman-Penrose formalism, for $D'\xi^\mu$ and $D'\alpha$.) This result would allow us to interchange the order of taking the $u$-derivative of $\eth^2\bar{\sigma}^o$ with the ``left'' $\eth$ to end up with $\eth\partial_u\eth\bar{\sigma}^o$ plus terms that eliminate each other upon integration over a compact surface. Since $\eth\partial_u\eth\bar{\sigma}^o$ itself would vanish when integrated, then the $u$-derivative of $\eth^2\bar{\sigma}^o$ does not contribute to the integration. [This $u$-derivative of $\eth^2\bar{\sigma}^o$ would appear in the $u$-derivative of the mass aspect, found in the $r^{-2}$ order of $\rho'$ --- see Eq. (\ref{yuiop}).]

Ergo, putting everything together into Eq. (\ref{finalint}), we find that all terms of orders $r^2$, $r$, $1$ and $r^{-1}$ exactly cancel out, giving:
\begin{align}
&-\oint{\left(\frac{\partial}{\partial u}(\Psi^o_2+\sigma^o\dot{\bar{\sigma}}^o)r^{-2}+O(r^{-3})\right)d^2A}\\
&=-\oint{\left(\left(|\dot{\sigma}^o|^2+\frac{\Lambda}{3}K|\sigma^o|^2+\frac{\Lambda}{3}|\eth'\sigma^o|^2+\frac{2\Lambda^2}{9}|\sigma^o|^4+\frac{\Lambda^2}{18}\textrm{Re}(\bar{\sigma}^o\Psi^o_0)\right)r^{-2}+O(r^{-3})\right)d^2A},
\end{align}
where terms involving an overall $\eth$ and $\eth'$ integrate to zero over the compact 2-surface that has no boundary, and integration by parts have been applied to the terms involving $\sigma^o\eth'\eth\bar{\sigma}^o$.

With $d^2A=r^2d^2S+O(r)$, where $d^2S$ is the area element of the topological 2-sphere of constant $u$ on $\mathcal{I}$, and then taking the limit as $r\rightarrow\infty$ so that $\Sigma_r$ becomes null infinity $\mathcal{I}$, we arrive at:
\begin{align}
&-\oint{\left(\frac{\partial}{\partial u}(\Psi^o_2+\sigma^o\dot{\bar{\sigma}}^o)\right)d^2S}\nonumber\\
&\phantom{-\oint{\left(\frac{\partial}{\partial u}\Psi^o_2\right)}}=-\oint{\left(|\dot{\sigma}^o|^2+\frac{\Lambda}{3}K|\sigma^o|^2+\frac{\Lambda}{3}|\eth'\sigma^o|^2+\frac{2\Lambda^2}{9}|\sigma^o|^4+\frac{\Lambda^2}{18}\textrm{Re}(\bar{\sigma}^o\Psi^o_0)\right)d^2S}.
\end{align}
One can then pull the $u$-derivative out of the surface integral [note that $\partial(d^2S)/\partial u=0$, with a proof given in Appendix \ref{appendix}] and produce the identical mass-loss formula as reported in Eq. (127) in Ref. \cite{Vee2016}, which we express here in terms of $M_B=-A^{-1}\oint{(\Psi^o_2+\sigma^o\dot{\bar{\sigma}}^o)d^2S}$ (the expression for the Bondi mass in the asymptotically flat case):
\begin{align}\label{masslossformulalah}
\frac{dM_B}{du}&=-\frac{1}{A}\oint{\left(|\dot{\sigma}^o|^2+\frac{\Lambda}{3}K|\sigma^o|^2+\frac{\Lambda}{3}|\eth'\sigma^o|^2+\frac{2\Lambda^2}{9}|\sigma^o|^4+\frac{\Lambda^2}{18}\textrm{Re}(\bar{\sigma}^o\Psi^o_0)\right)d^2S}.
\end{align}
Here, $A=4\pi$ is the area of the topological 2-sphere of constant $u$ on $\mathcal{I}$. Incidentally, the expression for ``$dL$'' of the Nester-Witten identity is \emph{precisely the rate of change of the Bondi mass for the asymptotically flat case, $dM_B/du$}.

\section{An identity for $K|\sigma^o|^2$; Quadrupole gravitational waves}\label{Gaussidentity}

\subsection{An identity for $K|\sigma^o|^2$}

Let $\eta(u,\theta,\phi)$ be a spin-weighted quantity with spin-weight $s=(p-q)/2$ [i.e. of type $(p,q)$], defined on $\mathcal{I}$ (so $\eta$ is independent of $r$). We can derive an identity for the Gauss curvature $K$ of the topological 2-sphere of constant $u$ on $\mathcal{I}$ using the commutator of the $\eth$ and $\eth'$ operators \cite{newpen65,Pen87}, as follows:
\begin{align}
|\eth\eta|^2-|\eth'\eta|^2&=\eth\eta\eth'\bar{\eta}-\eth'\eta\eth\bar{\eta}\\
&=\eth'(\bar{\eta}\eth\eta)-\eth(\bar{\eta}\eth'\eta)+\bar{\eta}[\eth,\eth']\eta\\
&=\eth'(\bar{\eta}\eth\eta)-\eth(\bar{\eta}\eth'\eta)+p\left(\rho\rho'-\sigma\sigma'+\Psi_2-\Phi_{11}-\frac{\Lambda}{6}\right)\eta\bar{\eta}\nonumber\\&\phantom{=}-q\left(\bar{\rho}\bar{\rho}'-\bar{\sigma}\bar{\sigma}'+\bar{\Psi}_2-\Phi_{11}-\frac{\Lambda}{6}\right)\eta\bar{\eta},
\end{align}
where we have used
\begin{align}
[\eth,\eth']&=-2i\textrm{Im}(\rho')\textrm{\th }+2i\textrm{Im}(\rho)\textrm{\th}'+p\left(\rho\rho'-\sigma\sigma'+\Psi_2-\Phi_{11}-\frac{\Lambda}{6}\right)\nonumber\\&\phantom{=}-q\left(\bar{\rho}\bar{\rho}'-\bar{\sigma}\bar{\sigma}'+\bar{\Psi}_2-\Phi_{11}-\frac{\Lambda}{6}\right)
\end{align}
[which is (4.14.1) from Ref. \cite{Pen87}] in the last line with the term $-2i\textrm{Im}(\rho')$\th\ being zero because \th\ is a partial derivative with respect to $r$ but $\eta$ is independent of $r$ (well, \th\ $=D=\partial/\partial r$ since $\gamma'=0$), and the term $2i\textrm{Im}(\rho)$\th$'$ being zero because $\rho$ is real. Since the left-hand side is manifestly real, then the right-hand side must also be real:
\begin{align}
|\eth\eta|^2-|\eth'\eta|^2&=\textrm{Re}(\eth'(\bar{\eta}\eth\eta)-\eth(\bar{\eta}\eth'\eta))+2s\textrm{Re}\left(\rho\rho'-\sigma\sigma'+\Psi_2-\Phi_{11}-\frac{\Lambda}{6}\right)\eta\bar{\eta}\\
&=\textrm{Re}(\eth'(\bar{\eta}\eth\eta)-\eth(\bar{\eta}\eth'\eta))-sK|\eta|^2,
\end{align}
where $2s=p-q$ and $-K=2\textrm{Re}(\rho\rho'-\sigma\sigma'+\Psi_2-\Phi_{11}-\Lambda/6)$. Integrating over a topological 2-sphere of constant $u$ on $\mathcal{I}$ (which has no boundary, so the terms with an overall $\eth$ or $\eth'$ derivative integrate to zero) gives the following identity:
\begin{align}
s\oint{K|\eta|^2d^2S}=\oint{|\eth'\eta|^2d^2S}-\oint{|\eth\eta|^2d^2S}.
\end{align}
For $\eta=\sigma^o$ which has spin-weight $s=2$, we have:
\begin{align}\label{Gaussidentityequation}
\frac{\Lambda}{3}\oint{K|\sigma^o|^2d^2S}=\frac{\Lambda}{6}\oint{|\eth'\sigma^o|^2d^2S}-\frac{\Lambda}{6}\oint{|\eth\sigma^o|^2d^2S}.
\end{align}

With this, we can replace the term involving the Gauss curvature in the mass-loss formula Eq. (\ref{masslossformulalah}) to obtain \footnote{Note that with this identity involving $K$, we did not actually need the explicit expression for $K$ given in Eq. (\ref{oneplussomething}), i.e. the mass-loss formula is independent of the choice of $\Theta(x^\mu)$.}:
\begin{align}\label{newmlf}
\frac{dM_B}{du}&=-\frac{1}{4\pi}\oint{\left(|\dot{\sigma}^o|^2+\frac{\Lambda}{2}|\eth'\sigma^o|^2-\frac{\Lambda}{6}|\eth\sigma^o|^2+\frac{2\Lambda^2}{9}|\sigma^o|^4+\frac{\Lambda^2}{18}\textrm{Re}(\bar{\sigma}^o\Psi^o_0)\right)d^2S}.
\end{align}

\subsection{Quadrupole gravitational waves}

In the absence of incoming radiation (so that $\Psi^o_0=0$), the mass-loss formula Eq. (\ref{newmlf}) reads:
\begin{align}\label{newmlfwithoutir}
\frac{dM_B}{du}&=-\frac{1}{4\pi}\oint{\left(|\dot{\sigma}^o|^2+\frac{\Lambda}{2}|\eth'\sigma^o|^2-\frac{\Lambda}{6}|\eth\sigma^o|^2+\frac{2\Lambda^2}{9}|\sigma^o|^4\right)d^2S}.
\end{align}
We see that for $\Lambda>0$, we have a \emph{negative-definite} term $|\eth\sigma^o|^2$, which is fine as long as the \emph{overall} expression on the right-hand side results in a mass-loss of $M_B$ such that the gravitational waves given off by an isolated system carries away a positive-definite energy.

Recall that the shear $\sigma^o$ is a spin-weighted quantity with spin-weight $s=2$, and $\eth\sigma^o$ has spin-weight $s=3$ since $\eth$ raises the spin-weight by one \cite{Pen87,Don}. When one studies a compact system in linearised theory, one can carry out a multipole expansion. (See for instance, Ref. \cite{Hobson} for a standard textbook treatment in the $\Lambda=0$ setup. For recent developments on the quadrupole formula with $\Lambda>0$, see Refs. \cite{ash3,gracos2,quadlambda}.) Spin-weighted quantities may be decomposed into spin-weighted spherical harmonics \cite{newpen65,swsh}. For a quantity with spin-weight $s$, it is expressible as a linear combination of these spin-weighted spherical harmonics, characterised by the non-negative integers $l$ with $l\geq|s|$, and integers $-l\leq m\leq l$. Hence in general, $\displaystyle\sigma^o(u,\theta,\phi)=P(u)\sum_{l=2}^\infty\sum_{m=-l}^l{_sY_{lm}}(\theta,\phi)$, where $s=2$. If the $\eth$ and $\eth'$ operators are defined on a round unit 2-sphere, then these $_sY_{lm}$ are the known standard spin-weighted spherical harmonics. But as we know when there is a non-zero cosmological constant $\Lambda\neq0$, these $_sY_{lm}$ correspond to those $\eth$ and $\eth'$ operators defined on the topological 2-sphere of constant $u$ on $\mathcal{I}$ [the definitions were given in Eqs. (\ref{eth}) and (\ref{ethprime})] \cite{Vee2016}. We are not attempting here to work out these $_sY_{lm}$ on the topological 2-sphere of constant $u$ on $\mathcal{I}$, as that deserves its own formal treatment in an exclusive study. Moreover, so far we only have the explicit metric for the axisymmetric case [see Eq. (\ref{metaxi})], but not the general one.

Instead, we note that in a multipole expansion of the compact source approximation, the quadrupole term corresponds to those where $l=2$, and the higher multipole terms are those with $l>2$. Now for \emph{purely quadrupole gravitational waves}, $\sigma^o$ would have only components with $l=2$. Since $\eth\sigma^o$ has spin-weight $s=3$, then $\eth\sigma^o=0$ because it can only have those components with $l\geq s=3$. We thus have from Eq. (\ref{newmlfwithoutir}) the following mass-loss formula:
\begin{align}\label{hohoho}
\frac{dM_B}{du}&=-\frac{1}{4\pi}\oint{\left(|\dot{\sigma}^o|^2+\frac{\Lambda}{2}|\eth'\sigma^o|^2+\frac{2\Lambda^2}{9}|\sigma^o|^4\right)d^2S},
\end{align}
which is \emph{manifestly} positive-definite with an overall negative sign for a universe with a positive cosmological constant, signifying that these \emph{purely quadrupole gravitational waves carry positive energy away from the isolated source}. This is in agreement with the linearised theory with $\Lambda>0$, as reported by Ashtekar, et al. \cite{ash3}.

With higher multipole terms however, there would necessarily be negative contributions to the expression for the energy carried away by the radiation [due to the $-|\eth\sigma^o|^2$ term in Eq. (\ref{newmlfwithoutir})]. Of course, what we need for a general $\sigma^o$ based on physical grounds would be that \emph{at least}:
\begin{align}\label{bound1}
\oint{\left(|\dot{\sigma}^o|^2+\frac{\Lambda}{2}|\eth'\sigma^o|^2-\frac{\Lambda}{6}|\eth\sigma^o|^2+\frac{2\Lambda^2}{9}|\sigma^o|^4\right)d^2S}\geq0,
\end{align}
where equality here must only be the case with $\sigma^o=0$. Alternatively, to ensure that the term involving the Gauss curvature $K$ itself is manifestly positive-definite, then from Eq. (\ref{Gaussidentityequation}), one can impose that:
\begin{align}\label{bound1point5}
|\eth'\sigma^o|\geq|\eth\sigma^o|
\end{align}
everywhere or only when integrated over $d^2S$. Another consideration would be to combine with the other $|\eth'\sigma^o|^2$ term in the mass-loss formula Eq. (\ref{masslossformulalah}) (which coincidentally carries the same factor of $\Lambda/3$ as that of the term with the Gauss curvature) to have this inequality:
\begin{align}\label{bound2}
\sqrt{3}|\eth'\sigma^o|\geq|\eth\sigma^o|
\end{align}
everywhere or only when integrated over $d^2S$. These provide some bounds for which the $l>2$ components of $\sigma^o$ would need to satisfy, to guarantee the positive-definiteness of the energy carried away by the outgoing gravitational waves.

\section{Concluding remarks}\label{conclusion}

In this paper, we have derived the mass-loss formula for an isolated gravitating system due to energy carried away by gravitational waves in a universe with a cosmological constant, using an integral formula on a hypersurface based on the Nester-Witten identity. The mass-loss formula was obtained by evaluating the integral formula using the asymptotic solutions for asymptotically de Sitter spacetimes near null infinity. Just like deriving the asymptotic solutions with $\Lambda\in\R$ \cite{Vee2016} being way more complicated than for $\Lambda=0$ \cite{newunti62}, the calculations involved here are much more tortuous when compared to the asymptotically flat case \cite{Fra97}. Nevertheless, we do eventually arrive at the identical mass-loss formula that was found in Ref. \cite{Vee2016} --- providing a pleasant consistency check that the Nester-Witten identity and the Bianchi identity return the same answer (as they should).

The Nester-Witten identity is written in the form ``$dL=S+E$'', where the ``$dL$'' is a differential of a 2-form ``$L$''. This provides a natural packaging of the many terms in the mass-loss formula Eq. (\ref{masslossformulalah}), where we have shown that the 3-form ``$dL$'' yields the usual expression for the rate of change of the Bondi mass $dM_B/du$ for any $\Lambda\in\R$. Whilst this is certainly not a physical justification for adopting such a generalisation of the Bondi mass to include a cosmological constant, we have backed up the choice of $M_\Lambda=M_B$ for any $\Lambda\in\R$ with a satisfactory result, viz. in a universe with $\Lambda>0$, the quadrupole gravitational waves would carry away a manifestly positive-definite energy from the isolated source. On top of that, we have also suggested some bounds [Eqs. (\ref{bound1})-(\ref{bound2})] that non-zero higher multipole contributions in $\sigma^o$ should satisfy (locally or integrated over $d^2S$) for $\Lambda>0$. This represents a step forward in the \emph{full non-linear theory of general relativity with a positive cosmological constant}.

The manifestly positive-definite result however, is \emph{false} for a universe with $\Lambda<0$. For instance, if $\Lambda$ is sufficiently small such that the $\Lambda^2$ term is negligible compared to the $\Lambda$ term and that $\sigma^o$ does not vary very strongly with $u$ such that $|\dot{\sigma}^o|^2$ is ignorable compared to $\Lambda|\eth'\sigma^o|^2/2$, then we see from Eq. (\ref{hohoho}) that the energy carried away by quadrupole gravitational waves in an anti-de Sitter-type universe is \emph{manifestly negative-definite}. In such a scenerio, perhaps one \emph{has to resort to moving that $\Lambda|\eth'\sigma^o|^2/2$ term to be absorbed into a new mass $M_\Lambda$, for $\Lambda<0$}, which suggests that we do indeed need to consider correction terms to the Bondi mass when $\Lambda\neq0$. From another perspective, this may be viewed as a theoretical basis for why a physically realistic universe cannot have a negative cosmological constant because even the lowest order (quadrupole) gravitational waves could carry away negative energy, as dictated by the full non-linear theory of general relativity.

\appendix

\section{The behaviour of empty asymptotically de Sitter spacetimes}\label{recapi}

In the asymptotically flat case, the assumption made by Newman and Penrose was only $\Psi^o_0=O(r^{-5})$, together with ``uniform smoothness'' \cite{newpen62}. With this single condition as their starting point, they were able to deduce the fall-offs for everything else, viz. the other dyad components of the Weyl spinor (the peeling property), spin coefficients and the unknown functions in the null tetrad. This was possible by harnessing the powerful mathematical result of Coddington and Levinson (see the details in Ref. \cite{newpen62}). Eventually, the freely-specifiable physical functions are:
\begin{align}
&\Psi_0,\textrm{on a null hypersurface }u=\textrm{constant}\\
&\Psi^o_1\textrm{ and Re}(\Psi^o_2),\textrm{on a }u=\textrm{constant cut of }\mathcal{I}\\
&\sigma^o,\textrm{on }\mathcal{I}.
\end{align}
Here, $u$ is a retarded null coordinate and so $u=$ constant defines an outgoing null hypersurface. The quantities $\Psi^o_1$, $\Psi^o_2$ and $\sigma^o$ are leading order terms of $\Psi_1$, $\Psi_2$, $\sigma$, respectively, when expanded in inverse powers of $r$ away from null infinity $\mathcal{I}$, where $r$ is an affine parameter of the outgoing null geodesics. Note that $\sigma^o$ determines $\Psi^o_3$ and $\Psi^o_4$ (leading order terms of $\Psi_3$ and $\Psi_4$, respectively).

For our study with $\Lambda\in\R$ in Ref. \cite{Vee2016} on the other hand, we shall not restrict ourselves to have a minimal set of assumptions. Instead, we assume from the start that all quantities are expressible as power series in inverse powers of $r$ of sufficiently many orders away from $\mathcal{I}$. This avoids mathematical technicalities in \emph{deriving} the fall-offs (which in our opinion, is unnecessary with regards to getting the \emph{physical} result, and moreover, it \emph{can} be deduced from admitting a smooth conformal compactifiability of the manifold). Here is our set of assumptions in Ref. \cite{Vee2016}:
\begin{enumerate}
	\item The fall-offs for the spin coefficients.
	\item The fall-offs for the functions in the null tetrad.
	\item The Weyl spinor vanishes on $\mathcal{I}$,
\end{enumerate}
as well as the fact that $\Psi_0=\Psi^o_0r^{-5}+O(r^{-6})$. Incidentally, it was much later realised that one does not need to assume $\Psi_0=O(r^{-5})$ when $\Lambda\neq0$, as this can be derived from the other assumptions (see Ref. \cite{Vee2017c}). Our ansatz for deriving the asymptotic solutions to the NP equations with $\Lambda\in\R$ are made by studying the Schwarzschild-de Sitter spacetime, and then calculating through all the 38 NP equations to find that they are consistent (as far as the relevant orders that we have carried this through), where these asymptotic solutions reduce exactly to the ones for $\Lambda=0$ as found in Ref. \cite{Pen88}. The peeling property of the Weyl (and Maxwell) spinor(s) with $\Lambda$ similarly follows from the same set of Bianchi identities (Maxwell equations) as in the $\Lambda=0$ case. For the extension to include Maxwell fields in Ref. \cite{Vee2017}, we correspondingly assume that the Maxwell spinor vanishes on $\mathcal{I}$ and $\phi_0=O(r^{-3})$ (but see Ref. \cite{Vee2017c} on how this fall-off for $\phi_0$ may not be required as an additional stipulation when $\Lambda\neq0$). The freely-specifiable physical functions with $\Lambda\neq0$ are the same as the $\Lambda=0$ setup.

Whilst our study and our ansatz are purely within the physical spacetime, they are all consistent with the fall-offs that were worked out by Szabados and Tod \cite{Szabados}. By assuming a smooth conformal compactifiability, they derived the fall-offs for the spin coefficients and the metric. One can carry out the appropriate boost transformation and null rotation to reconcile their null tetrad with ours here, to find that our set of assumptions agrees with their results --- including several gauge-fixing choices that lead to the vanishing of certain terms.

\section{Proof that the area element of the topological 2-spheres of constant $u$ on $\mathcal{I}$ is independent of $u$}\label{appendix}

We give a proof that the area of the topological 2-spheres of constant $u$ on $\mathcal{I}$ is the same as that of a round unit 2-sphere. Before dealing with the general case, we see that for the axisymmetric case, the metric is \cite{Vee2016}:
\begin{align}\label{metaxi}
g_{axi}=e^{2\Lambda f(u,\theta)}d\theta^2+e^{-2\Lambda f(u,\theta)}\sin^2{\theta}d\phi^2,
\end{align}
where $3f(u,\theta)=\int{\sigma^o(u,\theta)du}$. One can get the area element by calculating the square root of the determinant to find that $d^2S=\sin{\theta}d\theta d\phi$, which is just the usual area element of a round unit 2-sphere. This area element is independent of $u$, and the total area is just $A=\oint{d^2S}=\int_0^{2\pi}{\int_0^\pi{\sin{\theta}d\theta d\phi}}=4\pi$.

For the general case however, we do not have the explicit expression for the metric. Anyway, the area element for a topological 2-sphere of constant $u$ on $\mathcal{I}$ is:
\begin{align}
d^2S&=\frac{1}{r^2}d^2A,\textrm{ in the limit where }r\rightarrow\infty,
\end{align}
where $d^2A=-im_a\wedge\bar{m}_b$. In other words, $d^2A$ is the area element for the surfaces of constant $u$ and $r$. When expanded in inverse powers of $r$, its leading order term carries a factor of $r^2$, i.e. $d^2A=r^2d^2S+O(r)$. Division by $r^2$ and taking $r\rightarrow\infty$ would give the required area element on $\mathcal{I}$, which is $d^2S$.

Let us now evaluate the $u$-derivative of this area element $d^2S$. If one uses the Newman-Unti null tetrad which was employed in Refs. \cite{Vee2016,Vee2017}, then consider $\vec{Z}=\vec{n}_{\textrm{Saw}}-U\vec{l}_{\textrm{Saw}}=\vec{\partial}_u+X^\mu\vec{\partial}_\mu$ \footnote{On the other hand, if one uses the null tetrad that Frauendiener employed \cite{Fra97}, which is the one taken in this paper to evaluate his integral formula, then consider $\vec{Z}=\vec{n}-(U-\omega\bar{\omega})\vec{l}=\vec{\partial}_u+(X^\mu-2\textrm{Re}(\bar{\omega}\xi^\mu))\vec{\partial}_\mu$. Either way, one gets the same result --- see the next footnote.}. We can calculate the $u$-derivative of $d^2S$ by taking the Lie derivative of $d^2S$ along $\vec{\partial}_u$.

Note that $d^2S$ has no $r$ dependence. The Lie derivative of $d^2S$ along $X^\mu\vec{\partial}_\mu$ essentially unravels into taking derivatives of $d^2S$ with respect to the two angles (or the two coordinates intrinsic to this 2-surface). The factor of $X^{\mu}=O(r^{-3})$ [Eq. (\ref{Xmu})] would imply that in the limit where $r\rightarrow\infty$, this would be zero.

Therefore, we are left with working out the Lie derivative of $d^2S$ along $\vec{Z}$. With a calculation similar to that in Section \ref{Zzz}:
\begin{align}
\frac{\partial}{\partial u}(d^2S)&=\pounds_{\vec{Z}}(d^2S)\\
&=\frac{1}{r^2}\pounds_{\vec{Z}}(d^2A),\textrm{ in the limit where }r\rightarrow\infty\\
&=\frac{2}{r^2}\textrm{Re}(U\rho_{\textrm{Saw}}-\rho'_{\textrm{Saw}})d^2A,\textrm{ in the limit where }r\rightarrow\infty\\
&=2\textrm{Re}(U\rho_{\textrm{Saw}}-\rho'_{\textrm{Saw}})d^2S,\textrm{ in the limit where }r\rightarrow\infty.
\end{align}
Explicit calculations show that $2\textrm{Re}(U\rho_{\textrm{Saw}}-\rho'_{\textrm{Saw}})=O(r^{-2})$ \footnote{Similarly, $2\textrm{Re}((U-\omega\bar{\omega})\rho-\rho')=O(r^{-2})$, which is what appears in the corresponding calculation with the other null tetrad.}, so:
\begin{align}
\frac{\partial}{\partial u}(d^2S)=0,
\end{align}
i.e. the area element of the topological 2-sphere of constant $u$ on $\mathcal{I}$ is independent of $u$, and so the total area of such a topological 2-sphere is the same as that of the round unit 2-sphere, which is $4\pi$.

\begin{acknowledgments}
I wish to thank J{\"o}rg Frauendiener for the discussions. V.-L. Saw was supported by the University of Otago Doctoral Scholarship and the University of Otago Postgraduate Publishing Bursary.
\end{acknowledgments}

\bibliographystyle{spphys}       
\bibliography{Citation}

\end{document}